\documentclass[reprint,aps,pre,showpacs,showkeys]{revtex4-1}

\usepackage{natbib}
\usepackage{graphics}
\usepackage{bm}
\usepackage{hyperref}

\begin{document}

\title{Nonlinear normal modes in electrodynamic systems:
A nonperturbative approach}
\author{A.~V.~Kudrin}
\email{kud@rf.unn.ru}
\author{O.~A.~Kudrina}
\author{E.~Yu.~Petrov}
\affiliation{Department of Radiophysics, University of Nizhny Novgorod,
23 Gagarin Ave., Nizhny Novgorod 603950, Russia}

\begin{abstract}
We consider electromagnetic nonlinear
normal modes in cylindrical cavity resonators filled with
a nonlinear nondispersive medium. The key feature of the analysis is that
exact analytical solutions of the nonlinear field equations are employed to study the
mode properties in detail. Based on such a nonperturbative approach,
we  rigorously prove that the total energy of free nonlinear oscillations in a distributive conservative system,
such as that considered in our work, can exactly coincide with the sum of energies of the normal modes of the system.
This fact implies that the energy orthogonality property, which has so far been known to take place
only for linear oscillations and fields, can also be observed in the nonlinear oscillatory system.
\end{abstract}


\maketitle

\section{Introduction}
The concept of normal modes is a milestone in the theory of linear
oscillating systems and has had a significant impact on all fields of physics~\cite{Lan,Jac,Dub,Par}.
As is known, a linear normal mode (LNM) is a free sinusoidal oscillation in a
conservative dynamical system with constant parameters. In bounded
distributed dynamical systems, an infinite but countable
set of LNMs exists. In such systems, each LNM is characterized by its frequency and shape,
and satisfies homogeneous partial differential equations (PDEs) of motion
with given boundary conditions. A family of LNMs possesses the following
important properties, which allows one to solve a whole set of problems
related to the calculation of free and forced motions in a linear
system.

1. {\em Invariance}. Each LNM can be excited independently of other LNMs
by the specific choice of the initial conditions.

2. {\em Completeness}. An arbitrary oscillatory process in the system can
be expressed as a superposition of LNMs.

3. {\em Energy orthogonality}. The total energy present in the system due to
a free oscillatory process is the sum of the LNM energies.

Since nature is nonlinear, LNMs can only be regarded as very useful
mathematical models describing actual oscillations of nonlinear systems in
the weak-amplitude limit. However, the following question naturally arises:
Do strongly nonlinear systems admit such specific motions that their properties allow one to
consider them as nonlinear normal modes (NNMs), i.e., nonlinear generalizations
of the LNMs of the underlying linear systems? An affirmative answer
this question with respect to the lumped systems has been done in the
seminal works of Lyapunov~\cite{Lya} and Rosenberg~\cite{R60,R62,R66}. Rosenberg defined
an NNM as a vibration in unison of the mechanical system, i.e.,
a synchronous oscillation during which all the displacements of the
material points of the system reach their extreme values and pass through
zero simultaneously. A definition of the NNM in an autonomous distributed
system in terms of the dynamics on a two-dimensional invariant manifold
in phase space has been proposed by Shaw and Pierre~\cite{S94}. Using the invariant
manifold approach, they have also developed the technique of asymptotic series expansions
for constructing NNMs for a rather wide class of nonlinear
$1+1$D autonomous systems. In the past decades, the concept of NNMs
in mechanical systems has been studied extensively by a large number
of workers (see, e.g., works~\cite{Man,S93,V92,V96,Lac,Ker,Pee} and references therein). At the same time,
NNMs in electrodynamic systems remain poorly studied.

In this work, we present a nonperturbative approach to the concept
of NNMs in  an exactly integrable, nonlinear $2+1$D electrodynamic
system. It should be emphasized that all the forthcoming results are
exact, i.e., no asymptotic expansions will be used. Our approach
is based on the theory developed in~\cite{Kud,P10}. The nonlinear PDEs considered
in~\cite{Kud,P10} and herein depend explicitly on the independent variable (radial
coordinate) and can formally be regarded as $1+1$D nonautonomous systems.
Therefore, the approach proposed in~\cite{S94}, which is restricted to the
autonomous systems, cannot be applied directly. In the present study, we
define an NNM as follows: {\em The nonlinear normal mode (NNM) of a bounded
distributed conservative system is an oscillatory motion in which all of
the field quantities oscillate periodically in time with the same constant
period in the whole volume of the system}. Each of the NNM fields must exactly satisfy the nonlinear
PDEs of motion and the boundary conditions, and reduces to the
corresponding LNM field in the weak-field limit. Note that a similar
generalized definition of the NNM  as an unnecessarily
synchronous periodic motion in a mechanical system was proposed in~\cite{Pee}.

In what follows, we construct the electromagnetic NNMs in
cylindrical resonators filled with a nonlinear nondispersive medium and discuss
their properties in detail. It will be shown that the considered NNMs, as
their linear counterparts, exactly satisfy the first above-mentioned
property of invariance and, what is quite remarkable, the third (energy
orthogonality) property.

\section{Basic Equations}

Consider electromagnetic fields in a bounded cylindrical cavity of radius $a$ and height $L$. We assume
that the $z$ axis of a cylindrical coordinate system ($r$, $\phi$, $z$) is aligned with
the cavity axis and limit ourselves to consideration of axisymmetric field
oscillations, in which only the $E_{z}$ and $H_{\phi}$ components are nonzero. We will also
assume that the cavity is filled with a nonlinear nondispersive medium in which the longitudinal component of the electric displacement can be
represented as $D_{z}=D_{0}+\alpha^{-1}\epsilon_{0}\varepsilon_{1}[\exp(\alpha E_{z})-1]$, where $D_{0}$, $\varepsilon_{1}$, and $\alpha$ are
certain constants.
Since even powers of $E_z$
are present in the series expansion of $D_z$, the medium does not possess
a center of inversion. This is inherent in, e.g., uniaxial pyroelectric
and ferroelectric crystals, provided that the $z$ axis is aligned with
the crystallographic symmetry axis. With
appropriately chosen constants $D_{0}$, $\varepsilon_{1}$, and $\alpha$,
such an exponential constitutive relation correctly describes dielectric properties of actual media lacking a
center of inversion in the case of weak nonlinearity where we can
restrict ourselves to the quadratic (in $E_z$)
correction term to the linear dependence of $D_z$ on $E_z$ (see~\cite{Kud,P10} for more details).
Note that the model of a nonlinear distributed system presented herein and
its generalizations were also discussed in works~\cite{X10,Esk,X11,X12,P12}. In this case, the Maxwell
equations read
\begin{eqnarray}
{\partial H\over \partial r}+{1\over r}H=\varepsilon(E)\,{\partial E\over \partial t},\label{eq1}\\
{\partial E\over \partial r}=\mu_{0}\,{\partial H\over \partial t},\label{eq2}
\end{eqnarray}
where $E\equiv E_{z}(r,t)$, $H\equiv H_{\phi}(r,t)$, and
\begin{equation}
\varepsilon(E)\equiv dD_{z}/dE=\epsilon_{0}\varepsilon_{1}\exp(\alpha E).\label{eq3}
\end{equation}

An exact solution to the system of equations~(\ref{eq1}) and~(\ref{eq2}) can be written
in implicit form as~\cite{P10}
\begin{eqnarray}
\hspace{-5mm}
&&E={\cal E}\left(\rho\,e^{\alpha E/2},
\tau+\alpha Z_{0}\rho H/(2\varepsilon_{1}^{1/2})\right),\nonumber\\
\hspace{-5mm}
&&H=Z_{0}^{-1}{\varepsilon^{1/2}_{1}}\,e^{\alpha E/2}
{\cal H}\left(\rho\,e^{\alpha E/2},
\tau+\alpha Z_{0}\rho H/(2\varepsilon_{1}^{1/2})\right).\label{eq4}
\end{eqnarray}
Hereafter,  $\rho=r/a$, $\tau=t(\epsilon_{0}\varepsilon_{1}\mu_{0})^{-1/2}/a$, and
$Z_{0}=(\mu_{0}/\epsilon_{0})^{1/2}$. The functions
${\cal E}$ and ${\cal H}$ describe the electromagnetic field in a linear
medium and satisfy the equations
\begin{equation}
{\partial^{2} {\cal E}\over \partial \rho^2}+{1\over \rho}{\partial \cal E\over \partial \rho}={\partial^{2}{\cal E}\over \partial \tau^2}
\label{eq5}
\end{equation}
and
$${\partial \cal E\over \partial \rho}={\partial \cal H\over \partial \tau}.$$

The energy conservation law in the considered nonlinear medium can
easily be derived from the field equations~(\ref{eq1}) and~(\ref{eq2}). Multiplying~(\ref{eq1})
by $E$, and~(\ref{eq2}) by $H$, after some algebra we obtain
\begin{equation}
\frac{\partial w}{\partial t}+\nabla\cdot{\bf S}=0 \label{eq6}
\end{equation}
with the energy density
\begin{equation}
w=\epsilon_{0}\varepsilon_{1}\,
\frac{(\alpha E-1)\exp(\alpha E)+1}{\alpha^2}
+\mu_{0}\frac{H^{2}}{2} \label{eq7}
\end{equation}
and the Poynting vector
\begin{equation}
{\bf S}=-{\bf r}_{0}EH. \label{eq8}
\end{equation}
In the weak-field limit ($|\alpha E|\ll1$), expression~(7) reduces to
the well-known textbook formula $w=\epsilon_{0}\varepsilon_{1}E^{2}/2+\mu_{0}H^2/2.$

Based on formulas~(\ref{eq4}), the method proposed in~\cite{P10} makes
it possible to easily construct  exact periodic solutions of
the system of equations~(\ref{eq1}) and~(\ref{eq2}), starting from the LNMs which satisfy the
linear wave equation~(\ref{eq5}). It is the purpose of the forthcoming analysis to demonstrate that
the constructed solutions can actually be identified as NNMs of the considered electrodynamic systems.

\section{Nonlinear Normal Modes in a Cylindrical Cavity Resonator}

Analytical solution for the oscillations of the $E_{0n0}$ type in a
circular cylindrical cavity with perfectly conducting walls and
the nonlinear filling medium described by the dynamic permittivity~(\ref{eq3}) has been found in ~\cite{P10}
and is given by
\begin{eqnarray}
&&\hspace{-5mm}
E=A\,J_{0}(\kappa_{n}\,\rho\,e^{\alpha E/2})\,\cos(\kappa_{n}\theta),\nonumber\\
&&\hspace{-5mm}
H=-AZ^{-1}_{0}\varepsilon^{1/2}_{1}\,e^{\alpha E/2}
J_{1}(\kappa_{n}\,\rho\,e^{\alpha E/2})\,\sin(\kappa_{n}\theta),\label{eq9}
\end{eqnarray}
where $A$ is an arbitrary amplitude factor, $J_{m}$ is a Bessel function of the first kind of order $m$, $\kappa_{n}$ is the
$n$th positive root of the equation $J_{0}(\kappa)=0$, and $\theta=\tau+\alpha Z_{0}\rho H/(2\varepsilon_{1}^{1/2})$.
The electric and magnetic fields are described by the implicit functions
$E(r, t)$ and $H(r, t)$ which are solutions of system~(\ref{eq9}) of two transcendental
equations. These implicit functions exactly satisfy Maxwell equations (\ref{eq1}) and (\ref{eq2}), as well as the boundary conditions
\begin{equation}
E(a, t)=0,\quad |E(0, t)|<\infty. \label{eq10}
\end{equation}
The initial conditions can be obtained by substituting $\tau=0$ into formulas~(\ref{eq9}) to give
\begin{equation}
E=AJ_{0}(\kappa_{n}\rho e^{\alpha E/2}) \label{eq11}
\end{equation} at $t=0$ and
\begin{equation}
H(r, 0)\equiv0. \label{eq12}
\end{equation}
For a sufficiently large index $n$ such that $n>n^{*}(\alpha)$, where $n^{*}$ is a certain integer, the functions $E(r, t)$ and $H(r, t)$ become
ambiguous and solution~(\ref{eq9}), obtained without allowance for dispersion,
becomes inapplicable~\cite{P10}. For $n<n^{*}$, implicit functions $E$ and $H$ given by
formulas~(\ref{eq9}) describe continuous periodic oscillations with the time period $T_{n}=2\pi/\omega_{n}$,
where $\omega_{n}=\kappa_{n}(\epsilon_{0}\varepsilon_{1}\mu_{0})^{-1/2}a^{-1},$ satisfy the NNM definition formulated above, and
correspond to the NNMs of the $E_{0n0}$ (TM$_{0n0}$) type in the cavity. Specifying the integer index
$n$ and imposing the initial conditions~(\ref{eq11}) and~(\ref{eq12}), the motion of the
system follows the exact solution (\ref{eq9}) of the nonlinear boundary value
problem for equations~(\ref{eq1}) and~(\ref{eq2}), i.e., no oscillations with indices differing
from $n$ are excited. Hence, the considered NNMs satisfy the invariance
property.

Let us now consider some important features of NNMs, which
were not  pointed out in our previous work~\cite{P10}. First of all, the electric field
in these modes does not oscillate in unison in the whole cavity volume, i.e.,
the amplitudes of the field at different spatial points can reach their extreme
values and pass through zero at different instants of time. The same is true
for the magnetic field. It is clearly seen in Fig.~3 in~\cite{P10} that
there are no synchronous oscillations at different spatial points in the $E_{020}$
mode discussed therein. In this respect, the considered electromagnetic NNMs differ from
the well-known NNMs in lumped and $1+1$D distributed
mechanical systems~\cite{R66,S94}, and can be called ``internally resonant''~\cite{Ker,Pee}. It
will be shown below that the degree of  oscillation synchronism at different
points of the cavity resonator depends on the shape of the cavity.

We now calculate the total energy $W$ stored in each NNM of the $E_{0n0}$ type.
To this end, one should substitute the implicit functions $E$ and $H$ given by
formulas~(\ref{eq9}) into~(\ref{eq7}) and integrate $w$ over the cavity volume:
\begin{equation}
W=a^{2}\int\limits^{L}_{0}\int\limits^{2\pi}_{0}\int\limits^{1}_{0}w(\rho,\tau,\alpha)\rho d\rho d\phi
d z. \label{eq13}
\end{equation}
A remarkable result is that the quantity $W$ is independent of the
nonlinearity parameter $\alpha$ and exactly coincides with the total
energy $W_{0}^{(n)}$ of the corresponding linear $E_{0n0}$ mode in the cavity resonator
filled with  a linear medium that has the permittivity $\varepsilon=\epsilon_{0}\varepsilon_{1}={\rm const}$.
The rigorous proof of this fact is given in the Appendix.
Note that the quantity $W_{0}^{(n)}$ is calculated analytically as
\begin{eqnarray}
W_{0}^{(n)}&=&\pi\epsilon_{0}\varepsilon_{1}a^{2}L\int\limits_{0}^{1}({\cal E}^{2}+{\cal H}^{2})\rho d\rho=\nonumber \\
&=&\pi\epsilon_{0}\varepsilon_{1}a^{2}LA^{2}\left[\cos^{2}(\kappa_{n}\tau)\int\limits_{0}^{1}
J_{0}^{2}(\kappa_{n}\rho)\rho d\rho \right. + \nonumber \\
&&+\left. \sin^{2}(\kappa_{n}\tau)\int\limits_{0}^{1}J_{1}^{2}(\kappa_{n}\rho)\rho d\rho\right]= \nonumber \\
&=&\frac{\pi}{2}\epsilon_{0}\varepsilon_{1}a^{2}LA^{2}J_{1}^2(\kappa_{n}). \label{eq14}
\end{eqnarray}
It is worth also nothing that the fundamental frequencies $\omega_{n}$ of the NNMs
are independent of the field amplitude and the total energy, and coincide
with the eigenfrequencies of the LNMs of the $E_{0n0}$ type in the underlying linear system.
In the next section,  we will show that the above-mentioned notable features of the
NNMs hold for another electromagnetic system described by equations~(\ref{eq1}) and~(\ref{eq2}).

\section{Nonlinear Normal Modes in a Coaxial Resonator}

Assume that a coaxial cylindrical inner conductor of radius $b$ ($0<b<a$)
is inserted inside the cavity considered in the previous section. The NNMs of
the $E_{0n0}$ type in the nonlinear coaxial resonator can readily be constructed
using formulas~(\ref{eq4}) from the corresponding LNMs in the linear resonator
with a constant permittivity $\varepsilon=\epsilon_{0}\varepsilon_{1}$ ($\alpha=0$). The electric fields of the
LNMs of the $E_{0n0}$ type  must satisfy equation~(\ref{eq5}) and the following boundary
conditions on the perfectly conducting walls of the coaxial volume:
\begin{equation}
{\cal E}(b, t)={\cal E}(a, t)=0. \label{eq15}
\end{equation}
The LNMs fields are given by
\begin{eqnarray}
&&\hspace{-5mm}
{\cal E}=A\left[J_{0}(\mu_{n}\rho)Y_{0}(\mu_{n})-J_{0}(\mu_{n})Y_{0}(\mu_{n}\rho)\right]
\cos(\mu_{n}\tau),\nonumber \\
&&\hspace{-6mm}
{\cal H}=-A\left[J_{1}(\mu_{n}\rho)Y_{0}(\mu_{n})-J_{0}(\mu_{n})Y_{1}(\mu_{n}\rho)
\right]\sin(\mu_{n}\tau),\label{eq16}
\end{eqnarray}
where $Y_{m}$  is a Bessel function of the second kind of order $m$, and $\mu_{n}$ is
the $n$th positive root of the equation
\begin{equation}
J_{0}(\beta\mu)Y_{0}(\mu)-J_{0}(\mu)Y_{0}(\beta\mu)=0, \label{eq17}
\end{equation}
where $\beta=b/a$.

Substituting functions~(\ref{eq16}) into formulas~(\ref{eq4}), we obtain an exact
solution to the system of equations~(\ref{eq1}) and~(\ref{eq2}) in implicit form. It can be
verified that the boundary conditions~(\ref{eq15}) remain valid for the implicit
functions $E(r, t)$ and $H(r, t)$ given by~(\ref{eq4}) and~(\ref{eq16}). These implicit
functions are periodic in time with the period $T_{n}=2\pi/\omega_{n},$ where
$\omega_{n}=\mu_{n}(\epsilon_{0}\varepsilon_{1}\mu_{0})^{-1/2}a^{-1}$, since the transcendental equations~(\ref{eq4}) are
invariant with respect to the time shifts $\tau\rightarrow\tau+lT_{n}$ with integer $l$. Therefore,
equations~(\ref{eq4}) and~(\ref{eq16}) describe the fields of NNMs in the coaxial resonator with
perfectly conducting walls and a nonlinear nondispersive filling medium.

The fields of NNMs satisfy the initial conditions
\begin{equation}
E=A \left[J_{0}(\mu_{n}\rho e^{\alpha E/2})Y_{0}(\mu_{n})-J_{0}(\mu_{n})Y_{0}(\mu_{n}\rho e^{\alpha E/2})\right] \label{eq18}
\end{equation}
at $t=0$ and $H(r,0)\equiv 0$, and possess the invariance property.

\begin{figure}[ht]
\center
\includegraphics{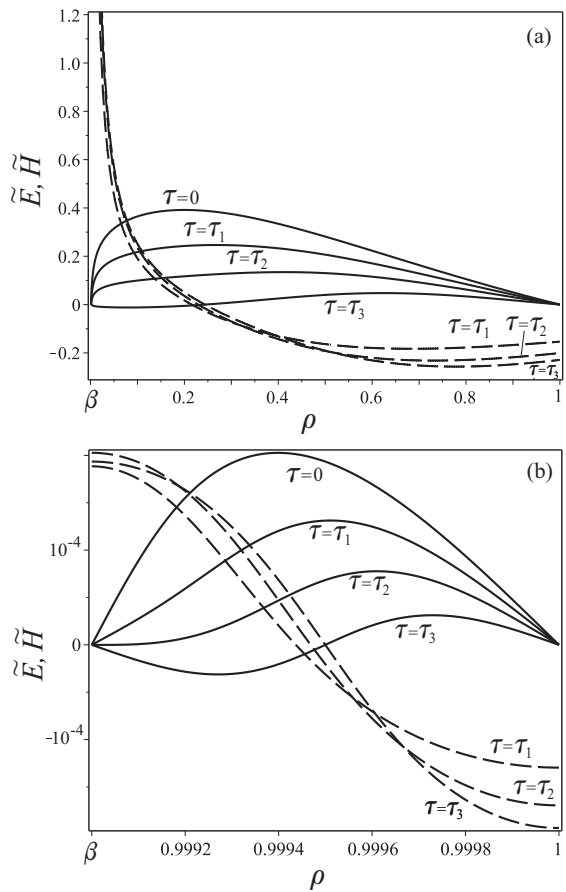}
\caption{Electric and magnetic fields as functions of $\rho$ (solid and dashed
lines, respectively) in
the $n=1$ mode of the coaxial resonator with $\beta=0.001$~(a) and $\beta=0.999$~(b)
at times $\tau=0$, $\tau_{1}=2\pi/(7\mu_{1})$, $\tau_{2}=2\pi/(5\mu_{1})$
and $\tau_{3}=\pi/(2\mu_{1})$}
\end{figure}

Let us turn to the results of some calculations by formulas~(\ref{eq4}) and~(\ref{eq16}).
Figure~1(a) shows the snapshots of the normalized electric and magnetic fields of the NNM of the $E_{010}$ type
in the coaxial resonator with $\beta=0.001$ ($n=1$ and $\mu_{1}=2{.}65...$) as functions
of $\rho$ at fixed instants of time~$\tau$. This case corresponds to a thin
inner conductor (coaxial wire) inside the cylindrical cavity. For comparison,
similar plots for the NNM of the $E_{010}$ type in the cavity with $\beta=0{.}999$ ($\mu_{1}=3141{.}59...$)
are presented in Fig.~1(b). In the limit $\beta\rightarrow1$, a coaxial geometry tends
to a thin flat layer. Note that Fig.~1 corresponds to the case of strong nonlinearity where
$\alpha A=1$.

The presented plots show that the electric fields at different spatial
points do not oscillate in unison. However, it is seen in Fig.~1 that the
degree of synchronism of the field in the  NNM of the $E_{010}$ type inside the cavity turns
out to be dependent on the value of $\beta$. The oscillations are closer to
synchronous ones for $\beta=0{.}001$.

The deviations of $E$ and $H$ from their values corresponding
to the $E_{010}$ mode in a resonator with $\varepsilon=\epsilon_{0}\varepsilon_{1}={\rm const}$
 $(\alpha=0)$ for $\beta=0.999$ are
more significant than those for $\beta=0{.}001$, i.e., the nonlinear effects become more
pronounced with increasing $\beta$. In the limit $\beta\rightarrow1$, the eigenvalue $\mu_{n}$ is close
to an integer multiple of $\pi$. This implies the more efficient interaction
of harmonics of the eigenfrequency in the spectrum of each NNM. One
may say that each NNM is more ``internally resonant'' in this case.

The total energy of each NNM of the $E_{0n0}$ type is again independent of
$\alpha$ and
exactly coincides with the total energy of the corresponding LNM in the
coaxial resonator filled with a linear medium (see the Appendix).

\section{More General Oscillations. Energy Orthogonality of Nonlinear Normal Modes}

In this section, we consider more general electromagnetic oscillations
which correspond to the presence of an infinite set of NNMs in a
cylindrical (noncoaxial) resonator. Let us state the following initial
and boundary conditions for the linear wave equation~(5):
\begin{eqnarray}
{\cal E}(\rho, 0)&=&\Phi(\rho),\quad \frac{\partial{\cal E}}{\partial \tau}(\rho, 0)=\Psi(\rho),\quad 0\leq\rho\leq1,\label{eq19}\\
{\cal E}(1, \tau)&=&0,\quad |{\cal E}(0, \tau)|<\infty,\quad 0<\tau<\infty, \label{eq20}
\end{eqnarray}
where $\Phi(\rho)$ and $\Psi(\rho)$ are given functions. The boundary value problem defined by (\ref{eq5}), (\ref{eq19}), and (\ref{eq20})
describes free electromagnetic oscillations with the given
initial field distribution in a cylindrical cavity specified by the relations
$\rho=r/a\leq1$ and $0\leq z\leq L,$ which is filled with a linear medium $(\alpha=0)$. The
solution to the linear boundary value problem specified by~(\ref{eq5}), (\ref{eq19}), and (\ref{eq20})
can be found in a standard way by the method of separation of variables~\cite{Cou}.
As a result, the functions ${\cal E}$ and ${\cal H}$ are written as
\begin{eqnarray}
&&\hspace{-7mm}
{\cal E}(\rho, \tau)=\sum_{n=1}^{\infty}J_{0}(\kappa_{n}\rho)\left[B_{n}\cos(\kappa_{n}\tau)+C_{n}\sin(\kappa_{n}\tau)\right], \nonumber \\
&&\hspace{-7mm}
{\cal H}(\rho, \tau)\!=\!-\sum_{n=1}^{\infty}\!J_{1}(\kappa_{n}\rho)\left[B_{n}\sin(\kappa_{n}\tau)
\!-\!C_{n}\cos(\kappa_{n}\tau)\right], \label{eq21}
\end{eqnarray}
where
\begin{eqnarray}
B_{n}=\frac{2}{J^{2}_{1}(\kappa_{n})}\int\limits_{0}^{1}\rho\Phi(\rho)J_{0}(\kappa_{n}\rho)
d\rho, \nonumber \\
C_{n}=\frac{2}{\kappa_{n}J^{2}_{1}(\kappa_{n})}\int\limits_{0}^{1}\rho\Psi(\rho)J_{0}(\kappa_{n}\rho)
d\rho. \label{eq22}
\end{eqnarray}
Substituting series~(\ref{eq21}) into formulas~(\ref{eq4}), we obtain an exact solution
to system of equations~(\ref{eq1}) and~(\ref{eq2}) in implicit form. Such an implicit solution describes
free electromagnetic oscillations which correspond to the presence of an
infinite set of NNMs in the nonlinear resonator. The implicit functions $E(\rho, \tau)$
and $H(\rho, \tau)$ determined by formulas~(\ref{eq4}) and~(\ref{eq21}) satisfy the boundary
conditions~(\ref{eq20}), but correspond to somewhat different initial conditions
compared with~(\ref{eq19}).

For example, let us specify the functions $\Phi$ and $\Psi$ in the simple form
\begin{eqnarray}
\Phi(\rho)&=&A(1-\rho^{2}),\label{eq23}\\
\Psi(\rho)&\equiv&0. \label{eq24}
\end{eqnarray}
This leads to
\begin{equation}
B_{n}=8A\kappa_{n}^{-3}[J_{1}(\kappa_{n})]^{-1},\quad  C_{n}=0. \label{eq25}
\end{equation}
At the initial time $\tau=0$, the electric field distribution $E(\rho, 0)$ in
the nonlinear resonator is defined by the transcendental equation
\begin{equation}
E=\sum_{n=1}^{\infty}B_{n}J_{0}(\kappa_{n}\rho e^{\alpha E/2}), \label{eq26}
\end{equation}
while the magnetic field $H\equiv 0$ as in the ``seeding''
linear problem (see~(\ref{eq24})).

\begin{figure}[ht]
\center
\includegraphics{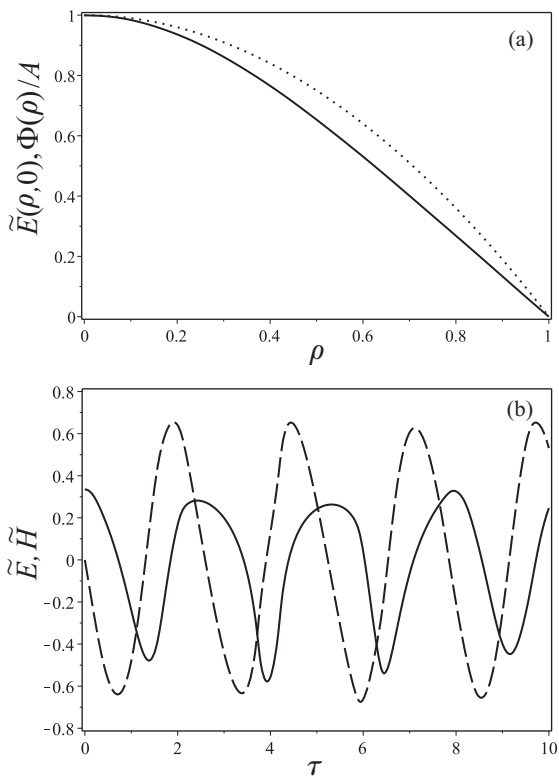}
\caption{(a)~Initial distribution of the electric field $E(\rho,0)/A$ (solid line)
and the function $\Phi(\rho)/A$ (dotted line). (b)~Oscillograms of the electric ($E$) and magnetic
($H$) fields of a nonlinear resonator at $\rho=0.75$ (solid
and dashed lines, respectively), calculated by formulas~(4) and (21) for $\alpha A=0.5$}
\end{figure}

The electric field distributions $\Phi(\rho)/A$ and $E(\rho, 0)/A$ in the linear and
nonlinear cases, respectively, are presented in Fig.~2(a). Figure~2(b) shows
oscillograms of the field components $E$ and $H$ at the fixed point $\rho=0.75$ for $\tau>0$.
Figure 2 was plotted for $\alpha A=0{.}5$.

It is important to note that the exact solution obtained for such $\alpha A$
corresponds to single-valued continuous functions $E$ and $H$ (see Fig.~2(b)).
While an NNM solution becomes ambiguous with increasing $n$ for any fixed $\alpha A$, the solution yielded by formulas~(\ref{eq4}) and~(\ref{eq21})
remains single-valued at moderate  $\alpha A$. This fact is stipulated by the rapidly
decreasing series coefficients $B_{n}$.

Despite the difference between the initial conditions given by~(\ref{eq23}) and ~(\ref{eq26}) (see Fig.~2(a)), the energy for the field distribution~(\ref{eq23}) in the linear cavity resonator
$(\alpha=0)$ coincides exactly with the energy of the field defined by~(\ref{eq26}) in
the nonlinear case where $\alpha\neq0$ (see the Appendix). Due to the energy conservation, the
same is true for an arbitrary time instant $\tau>0$. This fact implies a quite
remarkable energy orthogonality property of NNMs. The total energy $W$ of the complicated
oscillatory process, which is described by exact implicit solution~(\ref{eq4}), with
${\cal E}$ and ${\cal H}$ given by~(\ref{eq21}), is merely the sum of the NNM (or LNM)
energies~(\ref{eq14}):
\begin{eqnarray}
W\!&=&\!A^{-2}\sum_{n=1}^{\infty}B_{n}^{2}W_{0}^{(n)}
=\nonumber\\
\!&=&\!\pi\epsilon_{0}\varepsilon_{1}La^{2}\int\limits_{0}^{1}\Phi^{2}(\rho)\rho d\rho=\nonumber\\
\!&=&\!\frac{\pi}{6}\epsilon_{0}\varepsilon_{1}La^{2}A^{2}. \label{eq27}
\end{eqnarray}
The performed analysis shows that the energy orthogonality property
of the NNMs holds for a rather wide class of initial conditions~(\ref{eq19}). One should only ensure the convergence of Fourier series~(\ref{eq21})
and the absence of ambiguity of the implicit solutions.

For the electromagnetic fields governed by the nonlinear system of equations (1) and (2),
the principle of superposition does not hold and
the NNM fields lack the usual orthogonality property
which takes place for linear normal modes in resonators~\cite{Jac}. Moreover,
the interaction of the considered NNMs in the forced oscillations can result in
complex nonlinear dynamics with a singular-continuous (fractal) Fourier
spectrum~\cite{P12}. Therefore, the observed energy orthogonality property
of the NNMs seems especially interesting.

\section{Conclusions}

In this work, to the best of our knowledge, we have presented the first
nonperturbative approach
to the basic properties of NNMs in a distributed nonlinear system.
The approach does not require asymptotic expansions and
provides a rigorous theoretical formulation of the NNM properties.
It is to be emphasized that this formulation is not
restricted to consideration of weakly nonlinear systems.

In applying the developed approach, we have constructed exact
solutions to the electromagnetic fields of NNMs in cylindrical  resonators filled
with a nonlinear nondispersive medium. It has been shown that the field oscillations in the found NNMs
are periodic in time, but are not synchronous at different spatial points.
We have established that the total energy of any NNM is independent of
the nonlinearity parameter and exactly coincides with the energy of the
corresponding LNM in the linear resonator. We have also obtained an
exact solution which describes a more general oscillatory process
corresponding to the presence of a countable set of NNMs in the nonlinear
cylindrical resonator. Based on this solution, we have rigorously established
the energy orthogonality property of the NNM fields. A very intriguing and
physically important issue, which naturally arises from the present analysis
and still remains open, is whether energy orthogonality of NNMs is a
property inherent in the particular model of nonlinearity or can be extended to a wider
class of nonlinear distributed systems.

\begin{acknowledgments}
The theoretical study in this work was supported by the
Russian Science Foundation (Project No.~14--12--00510). Support for the numerical
calculations was provided through Contract Nos.~14.B25.31.0008 and
3.1252.2014/K from the Government of the
Russian Federation.
\end{acknowledgments}

\appendix*
\section*{Appendix}
\setcounter{equation}{0}

Let us show that the total energy of free nonlinear oscillations is
independent of the nonlinearity parameter $\alpha$ and coincides with the total energy
in the linear case $(\alpha=0)$. Due to the energy conservation in a cavity with
perfectly conducting walls and a nondispersive filling medium, it is sufficient
to prove this fact for an arbitrary fixed time instant (say, $\tau=0$). The electric
field distribution $E(\rho, 0)$ in the considered nonlinear oscillations is defined
by the transcendental equation
\begin{equation}
E={\cal E}(\rho e^{\alpha E/2}, 0), \label{A1}
\end{equation}
while $H(\rho, 0)\equiv0$. Introducing the notation $R=\rho\exp(\alpha E/2)$, we have
\begin{equation}
d R=(1+\alpha\rho E'_{\rho}/2)e^{\alpha E/2}d\rho. \label{A2}
\end{equation}
The derivative $E'_{\rho}$ can be found from~(\ref{A1}) as
\begin{equation}
E'_{\rho}=[1-\alpha e^{\alpha E/2}\rho {\cal E}'_{R}/2]^{-1}e^{\alpha E/2}{\cal E}'_{R}. \label{A3}
\end{equation}
Substituting~(\ref{A3}) into~(\ref{A2}) yields
\begin{equation}
d\rho=[1-\alpha R {\cal E}'_{R}/2]e^{-\alpha E/2}d R. \label{A4}
\end{equation}

Consider for clarity a cylindrical (noncoaxial)  resonator. Making the change
of variables and using~(\ref{A4}), from~(\ref{eq7}) and~(\ref{eq13})
one obtains the total energy
\begin{eqnarray}
W=&&2\pi\epsilon_{0}\varepsilon_{1}a^{2}L{\alpha}^{-2}\times \nonumber \\
&&\times\int\limits_{0}^{1}(\alpha E+e^{-\alpha E}-1)\times \nonumber \\
&&\times[1-\alpha R{\cal E}'_{R}/2]R d R. \label{A5}
\end{eqnarray}
Here, we have also taken into account the boundary condition $E=0$ at $\rho=1$,
implying that $R=1$ for $\rho=1$. It is convenient to rewrite~(\ref{A5}) as
\begin{eqnarray}
W=&&2\pi\epsilon_{0}\varepsilon_{1}a^{2}L\int\limits_{0}^{1}\left[-\frac{1}{2}{\cal E}{\cal E}'_{R}R^{2}+\right. \nonumber \\
&&+\frac{1}{2\alpha}{\cal E}'_{R}R^{2}+\frac{1}{\alpha^{2}}(e^{-\alpha E}-1)R + \nonumber \\
&&+\left.\frac{1}{\alpha}{\cal E}R-\frac{1}{2\alpha}{\cal E}'_{R}R^{2}e^{-\alpha E}\right]
d R. \label{A6}
\end{eqnarray}
Integrating the first term in the integrand of~(\ref{A6}) by parts and using the boundary condition
${\cal E}=0$ at $R=1$, we have
\begin{equation}
-\frac{1}{2}\int\limits_{0}^{1}{\cal E}{\cal E}'_{R}R^{2}d R={1\over 2}\int\limits_{0}^{1}{\cal E}^{2}
Rd R. \label{A7}
\end{equation}
Integrating the second and third terms in the integrand of~(\ref{A6}) by parts, one
can find that the result of integration cancels the last two terms in this integrand.
Finally, we get
\begin{equation}
W=\pi\epsilon_{0}\varepsilon_{1}a^{2}L\int\limits_{0}^{1}{\cal E}^{2}Rd R, \label{A8}
\end{equation}
which coincides with the total energy of the linear oscillations. For a coaxial resonator, the proof is similar.

In addition, it should be noted
that the total energy of the radially localized field distributions vanishing for $\rho\to \infty$  in an unbounded nonlinear medium, which can be the case for, e.g., cylindrical electromagnetic waves~\cite{P10}, is also
independent of the nonlinearity parameter $\alpha$.

\bibliography{Kudrin_Kudrina_Petrov_bibl}

\end{document}